# Near-field magneto-caloritronic nanoscopy on ferromagnetic nanostructures

E. Pfitzner,[2] X. Hu,[1] H. W. Schumacher,[1] A. Hoehl,[1] D. Venkateshvaran,[3] M. Cubukcu,[3] J.-W. Liao,[3] S. Auffret,[4] J. Heberle,[2] J. Wunderlich,[5,6] and B. Kästner[1]

[1]*Physikalisch-Technische Bundesanstalt, 38116 Braunschweig and 10587 Berlin, Germany*

[2]*Experimental Molecular Biophysics, Department of Physics, Freie Universität Berlin, 14195 Berlin, Germany*

[3]*Cavendish Laboratory, University of Cambridge, J. J. Thomson Avenue, Cambridge CB3 0HE, UK*

[4]*SPINTEC, CEA, CNRS and Université Grenoble Alpes, 17 rue des Martyrs, 38054, Grenoble, France*

[5]*Institute of Physics ASCR, v.v.i.; 162 53 Praha 6, Czech Republic*

[6]*Hitachi Cambridge Laboratory, Cambridge CB3 0HE, United Kingdom*

Near-field optical microscopy by means of infrared photocurrent mapping has rapidly developed in recent years. In this letter we introduce a near-field induced contrast mechanism arising when a conducting surface, exhibiting a magnetic moment, is exposed to a nanoscale heat source. The magneto-caloritronic response of the sample to near-field excitation of a localized thermal gradient leads to a contrast determined by the local state of magnetization. By comparing the measured electric response of a magnetic reference sample with numerical simulations we derive an estimate of the field enhancement and the corresponding temperature profile induced on the sample surface.

Scattering-type scanning near-field optical microscopy (*s*-SNOM) [1–3] has developed over the last decade into a powerful tool for the characterization of optical phenomena at the nanoscale. *s*-SNOM realizes sub-diffraction imaging and spectroscopy [4–10] and readily determines the topography [9,11,12], the mechanical phase [13], or the electrical response to optical near-field excitation [14,15]. Based on an atomic force microscope (AFM) *s*-SNOM utilizes a metal-coated tip brought in close proximity to the samples surface. When light is focused on the AFM probe, the tip acts as an optical antenna which strongly confines the incident electric field around the apex, thus, providing a nanoscale light source. Detecting the light scattered from the tip provides direct access to the optical material parameter [16], from which the chemical composition, electronic transport coefficients, or the mechanical strain can be extracted. The strongly confined electric near-field also acts as a thermal point source [17,18], lifting the diffraction limit present in focused laser heating [19] and driving local thermo-currents to be measured by external electrical contacts [17–24]. This method, also termed photocurrent nanoscopy, allows electrical transport properties to be investigated at nanoscale spatial resolution.

In this work we apply photocurrent nanoscopy to ferromagnetic nanostructures. In particular, we detect the electrical current which is generated by the thermal gradient localized in close proximity to the scanning tip illuminated by infrared (IR) radiation. We analyze the magneto-caloritronic contributions [25] which depend on the local magnetization distribution. The nanostructure we investigate is magnetized perpendicularly to the surface allowing us to image the local magnetization distribution by exploiting the anomalous Nernst effect (ANE) [26] and the anisotropic magneto-Seebeck effect [27]. In contrast to high-resolution scanning magnetic force microscopy where the sample magnetization can be affected by the stray-field of the scanning magnetic tip [28,29], our non-invasive magnetic photocurrent nanoscopy does not rely on the magnetic dipole interaction.

For tip-enhanced magneto-caloritronic nanoscopy an AFM (NanoWizard II, JPK Instruments, Germany) operated in tapping mode was used as shown schematically in Figure 1 (A). An Au coated Si cantilever (4XC-GG, NanoAndMore GmbH, Germany) with typical tip diameter below 30 nm oscillates at an amplitude $\Delta z = 50$ nm just above the sample surface at its mechanical resonance frequency $\Omega \sim 150$ kHz. The emission of a quantum cascade laser (QCL, $\sim 50$ mW at 1661 cm$^{-1}$, DRS Daylight Solutions Inc., CA, USA) was focused to the tip apex by a 90° off-axis parabolic mirror (diameter: 12.7 mm, focal length: 15 mm, angle-of-incidence: 75°). The IR induced temperature gradient, $\nabla T$, is indicated by the false color profile below the AFM tip in Figure 1 (A). The tip-mediated electric response of the sample to IR excitation was analyzed using a lock-in scheme. In short, the thermo-current generated in the magnetic wire was first amplified by a transimpedance amplifier (10$^6$ V/A, DHPCA-100, FEMTO Messtechnik GmbH, Germany) and further analyzed by a lock-in amplifier (HF2LI, Zurich Instruments, Switzerland) at the tip modulation frequency $\Omega$. Both the in-phase and out-of-phase components were registered while scanning the magnetic wire relative to the tip. The in-phase component typically exhibited a stronger contrast. The resulting thermal electromotive force (EMF), $V_T$, induced by the tip-enhanced IR radiation will be analyzed in the following as a function of the magnetization state of a ferromagnetic microbar.

In our experiment, we investigate the magnetization distribution in a 1μm wide and 60 μm long magnetic bar containing a central 500 nm wide triangular shaped notch (Figure 2 (A)). The microbar was defined by electron beam lithography on a poly(methyl methacrylate) (PMMA) resist layer. Subsequently, a Ta(3 nm)/Pt (3 nm)/Co(0.6 nm)/AlO$_x$(2 nm) magnetic multilayer was deposited on a thermally oxidized silicon wafer by DC magnetron sputtering followed by a lift-off procedure. The magnetic parameters in our Pt/Co/AlO$_x$ multilayers are as follows: exchange stiffness $A \cong 16$ pJ/m, saturation magnetization $M_s \cong 1.1$ MA/m, perpendicular anisotropy $K \cong 1.3$ MJ/m$^3$ and Dzyaloshinskii-Moriya interaction (DMI) parameter $D \cong 2.6$ mJ/m$^2$ [30,31]. The constriction is designed to act as a magnetic domain wall pinning center [32]. The bar is characterized by a perpendicular magnetic anisotropy and large interfacial DMI forcing magnetic domain walls to follow a Néel-like geometry with the magnetization direction at the domain wall center oriented along the bar direction [33].

The thermal EMF, $V_T$, was mapped over the microbar, as shown in Figure 2(B). The bar initially contained two oppositely oriented magnetic domains with the domain wall pinned at the constriction. The magnetization points along the $z$-direction ($M = M_s \mathbf{e}_z$) to the right of the constriction and 180° rotated ($M = -M_s \mathbf{e}_z$) to the left of the constriction. The corresponding $V_T$-map in Figure 2 (B) shows a gradient of $V_T$ along the $x$-direction near the constriction in the center of the wire. Moreover, the gradient of $V_T$ along the $y$-direction changes sign between the left and right hand side of the constriction. In order to understand the origin of the different contributions we first consider the local electric field $\mathbf{E}$ generated by the temperature gradient, $\nabla T$. In a coordinate system as shown in Figure 1 (B), $\varphi = 0$ is considered since the magnetization lies in the $x$-$z$-plane in the Néel-like domain wall. In this case, the local electric field $\mathbf{E}$ is given by

$$\begin{pmatrix} E_x \\ E_y \\ E_z \end{pmatrix} = \begin{pmatrix} S_\parallel \sin^2\theta + S_\perp \cos^2\theta & -S_N \cos\theta & (S_\parallel - S_\perp)\cos\theta \sin\theta \\ S_N \cos\theta & S_\perp & -S_N \sin\theta \\ (S_\parallel - S_\perp)\cos\theta\sin\theta & S_N \cos\theta & S_\parallel \cos^2\theta + S_\perp \sin^2\theta \end{pmatrix} \begin{pmatrix} [\nabla T]_x \\ [\nabla T]_y \\ [\nabla T]_z \end{pmatrix}, \quad (1)$$

where the anisotropic magneto-Seebeck coefficient $S_\parallel$ is measured when the temperature gradient is parallel to the magnetization while $S_\perp$ is measured when it is perpendicular to the magnetization direction. The elements $S_N$ represent the anomalous Nernst effect which can be estimated [28] by $S_N = |N_{ANE}| \mu_0 M_S$, with $\mu_0 M_S = 1.38$ T the magnetic moment of Co [34].

Our experimental setup is designed to detect the thermal EMF, $V_T$, between the two terminals along $x$. We estimate the thermal EMF, $dV_T = -E_x dx$, by integration along the microbar of width $w$ assuming two independent magnetic domains A ($-l/2 < x < 0$) and B ($l/2 > x > 0$), which are magnetized along the $z$-direction, i.e. $\theta = 0$ or $\pi$, and separated by a domain wall located a x = 0. For our sample with a high perpendicular anisotropy, the width of the Néel-like domain wall is less than 10 nm and hence too small to be resolved by our measurements. Therefore, in the domain wall region where $\theta$ is different from 0 or $\pi$ the contribution to $V_T$ generated by the anisotropic magneto-Seebeck effect could be neglected. We approximate the thermal EMF by the following formula:

$$V_T(x_0, y_0) = \int_{-l/2}^{l/2} dx \frac{1}{w(x)} \int_{-w(x)/2}^{w(x)/2} dy \left\{ -S_\perp [\nabla T(x,y,x_0,y_0)]_x + \cos\theta(x) S_N [\nabla T(x,y,x_0,y_0)]_y \right\}. \quad (2)$$

The position of the AFM tip is denoted by ($x_0$, $y_0$). In our uniaxial thin film samples, the measured thermal EMF depends predominantly on the local perp.-to-plane saturation magnetization via $\theta$ for the two domains.

The contributions to the thermal EMF compensate as long as the temperature variation due to the thermal point source falls off completely within the microbar and within one domain. If the thermal point source approaches the constriction, the origin of the measured contrast is dominated by the term $S_\perp [\nabla T]_x$, i.e. due to the uncompensated Seebeck effect contributions along the bar towards the notch. The thermal gradient in the term $\cos \theta(x) S_N [\nabla T]_y$, i.e. perpendicular to the bar and close to the edges of the bar gives rise to an additional thermal EMF contribution originating from the ANE since here the thermal gradient perpendicular to the bar varies asymmetrically and changes sign at opposite edges.

In order to extract the odd-under-magnetization reversal ANE contribution, the $V_T$ – map of the bar magnetized homogeneously in the $-z$-direction (single domain) was measured and subtracted from the reverse magnetized case. The difference of the thermal EMF is plotted in Figure 2 (C). Since the contributions due to $S_\perp [\nabla T]_x$ do not change when the magnetization is switched from $\boldsymbol{M} = -M_s \, \mathbf{e}_z$ to $\boldsymbol{M} = M_s \, \mathbf{e}_z$, they cancel each other while the ANE contributions double. Accordingly, the generation of thermal EMF near the edges can be seen along the whole bar, with opposite sign on either side, following the sign change of the $y$-component of $\nabla T$. The topography-induced artifacts at the edge of the bar displayed in Figure 2 (B) are drastically reduced correspondingly upon subtraction, since they are as well not sensitive to the reversed magnetization of the bar itself. Similarly, in Figure 2 (D) the domain wall location can be visualized by subtracting the homogeneously magnetized map from the $V_T$ – map in Figure 2 (B). However, the signal-to-noise ratio in the present data does not allow quantifying the lateral size of the domain wall.

For a semi-quantitative analysis, the following considers a $V_T$ – trace along the $y$-direction sufficiently far away from the constriction for the two magnetization directions, as shown in Figure 3 (A). The trace has been averaged over 12 neighboring lines with $\Delta x = 15$ nm spacing and subsequently smoothed by a Savitzky-Golay filter. The inversion of $V_T$ upon magnetization reversal is verified. A line scan without illumination by the QCL, but otherwise identical experimental conditions, didn't yield the characteristic asymmetric shape (see supplementary information). We also simulated the temperature distribution caused by the illuminated tip using a circularly shaped heat source. A Gaussian power density distribution of 50 nm in diameter (FWHM) was assumed, where the peak value serves as fitting parameter. With dedicated heater structures (not shown) on this particular sample we were able to determine [27] the ANE coefficient for our microbar experimentally as $|N_{ANE}| = 0.054$ µV/KT, from which we obtain the trace $V_T(y)$ in Figure 3 (B) by employing Eqn. (2). It reproduces the anti-symmetric shape and absolute range of variation of the measured $V_T$ when a peak power density loss at the surface of 4 GW/m² (4 mW/µm²) was assumed, with an estimated input power density close to the tip of 0.01 GW/m². This is consistent with a field enhancement factor of about 20 - 30 as expected for metallized AFM tips [35]. The inset of Figure 3 (B) also shows the corresponding temperature distribution, indicating a temperature rise of 20 - 30K of the surface underneath the tip, which is still well below the Curie temperature of our thin Co layer. The ability to estimate the local temperature is an important byproduct of our measurement.

In summary, the magneto-caloritronic response of a conducting sample to near-field excitation leads to a novel contrast mechanism at magnetic domain boundaries as well as near the edges of the magnetic nanostructure due to the anomalous Nernst effect. The contrast was demonstrated by reversing the magnetization of the nanostructure resulting in a corresponding reversal of the ANE generated thermal EMF. The interpretation was supported by a 2D numerical simulation. Magneto-caloritronic nanoscopy can provide information on magnetic surface properties without relying on the magnetic dipole interaction.

**Acknowledgments**

This work was supported by the Deutsche Forschungsgemeinschaft through grant HE 2063/5-1 to JH. The work also received funding from the ERC synergy grant No. 610115.

**References**

1. B. Knoll and F. Keilmann, "Infrared conductivity mapping for nanoelectronics," Appl. Phys. Lett. **77**, 3980 (2000).
2. F. Keilmann and R. Hillenbrand, "Near-field nanoscopy by elastic light scattering from a tip," in *Nano-Optics and Near-Field Optical Microscopy*, A. Zayats and D. Richard, eds. (Artech House, Boston, London, 2009), pp. 235–265.
3. A. Hartschuh, "Tip-enhanced near-field optical microscopy," Angew. Chemie - Int. Ed. **47**, 8178–8191 (2008).
4. A. C. Jones and M. B. Raschke, "Thermal Infrared Near-Field Spectroscopy," Nano Lett. **12**, 1475–1481 (2012).
5. F. Huth, A. Govyadinov, S. Amarie, W. Nuansing, F. Keilmann, and R. Hillenbrand, "Nano-FTIR absorption spectroscopy of molecular fingerprints at 20 nm spatial resolution.," Nano Lett. **12**, 3973–8 (2012).
6. P. Hermann, A. Hoehl, P. Patoka, F. Huth, E. Rühl, and G. Ulm, "Near-field imaging and nano-Fourier-transform infrared spectroscopy using broadband synchrotron radiation.," Opt. Express **21**, 2913–9 (2013).
7. H. A. Bechtel, E. A. Muller, R. L. Olmon, M. C. Martin, and M. B. Raschke, "Ultrabroadband infrared nanospectroscopic imaging.," Proc. Natl. Acad. Sci. U. S. A. **111**, 7191–7196 (2014).
8. B. T. O'Callahan, W. E. Lewis, S. Möbius, J. C. Stanley, E. a. Muller, and M. B. Raschke, "Broadband infrared

vibrational nano-spectroscopy using thermal blackbody radiation," Opt. Express **23**, 32063 (2015).

9. B. Pollard, F. C. B. Maia, M. B. Raschke, and R. O. Freitas, "Infrared Vibrational Nanospectroscopy by Self-Referenced Interferometry," Nano Lett. **16**, 55–61 (2016).

10. I. Amenabar, S. Poly, W. Nuansing, E. H. Hubrich, A. A. Govyadinov, F. Huth, R. Krutokhvostov, L. Zhang, M. Knez, J. Heberle, A. M. Bittner, and R. Hillenbrand, "Structural analysis and mapping of individual protein complexes by infrared nanospectroscopy.," Nat. Commun. **4**, 2890 (2013).

11. I. Amenabar, S. Poly, M. Goikoetxea, W. Nuansing, P. Lasch, and R. Hillenbrand, "Hyperspectral infrared nanoimaging of organic samples based on Fourier transform infrared nanospectroscopy," Nat. Commun. **8**, 14402 (2017).

12. S. Berweger, D. M. Nguyen, E. a Muller, H. a Bechtel, T. T. Perkins, and M. B. Raschke, "Nano-chemical infrared imaging of membrane proteins in lipid bilayers.," J. Am. Chem. Soc. **135**, 18292–5 (2013).

13. B. Kästner, C. M. Johnson, P. Hermann, M. Kruskopf, K. Pierz, A. Hoehl, A. Hornemann, G. Ulrich, J. Fehmel, P. Patoka, E. Rühl, and G. Ulm, "Infrared nanospectroscopy of phospholipid and surfactin monolayer domains," ACS Omega (2018).

14. A. Woessner, M. B. Lundeberg, Y. Gao, A. Principi, P. Alonso-González, M. Carrega, K. Watanabe, T. Taniguchi, G. Vignale, M. Polini, J. Hone, R. Hillenbrand, and F. H. L. Koppens, "Highly confined low-loss plasmons in graphene–boron nitride heterostructures," Nat. Mater. 1–5 (2014).

15. A. Woessner, P. Alonso-gonzalez, M. B. Lundeberg, G. Navickaite, Y. Gao, Q. Ma, D. Janner, K. Watanabe, T. Taniguchi, V. Pruneri, P. Jarillo-herrero, J. Hone, R. Hillenbrand, and F. H. L. Koppens, "High resolution near-field photocurrent measurements reveal optoelectronic properties of graphene," in *Graphene 2015* (2014), Vol. 487, p. 5094.

16. E. A. Muller, B. Pollard, and M. B. Raschke, "Infrared Chemical Nano-Imaging: Accessing Structure, Coupling, and Dynamics on Molecular Length Scales," J. Phys. Chem. Lett. **6**, 1275–1284 (2015).

17. A. Woessner, P. Alonso-González, M. B. Lundeberg, Y. Gao, J. E. Barrios-Vargas, G. Navickaite, Q. Ma, D. Janner, K. Watanabe, A. W. Cummings, T. Taniguchi, V. Pruneri, S. Roche, P. Jarillo-Herrero, J. Hone, R. Hillenbrand, and F. H. L. Koppens, "Near-field photocurrent nanoscopy on bare and encapsulated graphene," Nat. Commun. **7**, 10783 (2016).

18. J. C. Karsch, J. M. Bartell, and G. D. Fuchs, "Near-field coupling of gold plasmonic antennas for sub-100 nm magneto-thermal microscopy," APL Photonics **2**, 086104 (2017).

19. J. M. Bartell, D. H. Ngai, Z. Leng, and G. D. Fuchs, "Towards a table-top microscope for nanoscale magnetic imaging using picosecond thermal gradients," Nat. Commun. **6**, 8460 (2015).

20. J. W. P. Hsu, E. A. Fitzgerald, Y. H. Xie, and P. J. Silverman, "Near-field scanning optical microscopy imaging of individual threading dislocations on relaxed GexSi1−x films," Appl. Phys. Lett. **65**, 344–346 (1994).

21. S. K. Buratto, J. W. P. Hsu, J. K. Trautman, E. Betzig, R. B. Bylsma, C. C. Bahr, and M. J. Cardillo, "Imaging InGaAsP quantum-well lasers using near-field scanning optical microscopy," J. Appl. Phys. **76**, 7720–7725 (1994).

22. C. R. McNeill, H. Frohne, J. L. Holdsworth, J. E. Furst, B. V King, and P. C. Dastoor, "Direct Photocurrent Mapping of Organic Solar Cells Using a Near-Field Scanning Optical Microscope," Nano Lett. **4**, 219–223 (2004).

23. Y. Gu, E.-S. Kwak, J. L. Lensch, J. E. Allen, T. W. Odom, and L. J. Lauhon, "Near-field scanning photocurrent microscopy of a nanowire photodetector," Appl. Phys. Lett. **87**, 43111 (2005).

24. T. Mueller, F. Xia, M. Freitag, J. Tsang, and P. Avouris, "Role of contacts in graphene transistors: A scanning photocurrent study," Phys. Rev. B **79**, 245430 (2009).

25. G. E. W. Bauer, E. Saitoh, and B. J. van Wees, "Spin caloritronics," Nat. Mater. **11**, 391 (2012).

26. A. v. Ettingshausen and W. Nernst, "Ueber das Auftreten electromotorischer Kräfte in Metallplatten, welche von einem Wärmestrome durchflossen werden und sich im magnetischen Felde befinden," Ann. Phys. **265**, 343–347 (1886).

27. P. Krzysteczko, X. Hu, N. Liebing, S. Sievers, and H. W. Schumacher, "Domain wall magneto-Seebeck effect," Phys. Rev. B **92**, 140405 (2015).

28. P. Krzysteczko, J. Wells, A. Fernández Scarioni, Z. Soban, T. Janda, X. Hu, V. Saidl, R. P. Campion, R. Mansell, J.-H. Lee, R. P. Cowburn, P. Nemec, O. Kazakova, J. Wunderlich, and H. W. Schumacher, "Nanoscale thermoelectrical detection of magnetic domain wall propagation," Phys. Rev. B **95**, 220410 (2017).

29. M. Cubukcu, D. Venkateshvaran, A. Wittmann, S.-J. Wang, R. Di Pietro, S. Auffret, L. Vila, J. Wunderlich, and H. Sirringhaus, "Electrical nucleation and detection of single 360° homochiral Néel domain walls measured using the anomalous Nernst effect," Appl. Phys. Lett. **112**, 262409 (2018).

30. J.-P. Tetienne, T. Hingant, L. J. Martínez, S. Rohart, A. Thiaville, L. H. Diez, K. Garcia, J.-P. Adam, J.-V. Kim, J.-F. Roch, I. M. Miron, G. Gaudin, L. Vila, B. Ocker, D. Ravelosona, and V. Jacques, "The nature of domain walls in ultrathin ferromagnets revealed by scanning nanomagnetometry," Nat. Commun. **6**, 6733 (2015).

31. S. Pizzini, J. Vogel, S. Rohart, L. D. Buda-Prejbeanu, E. Jué, O. Boulle, I. M. Miron, C. K. Safeer, S. Auffret, G. Gaudin, and A. Thiaville, "Chirality-Induced Asymmetric Magnetic Nucleation in Pt/Co/AlOx Ultrathin

Microstructures," Phys. Rev. Lett. **113**, 47203 (2014).

32. T. Koyama, G. Yamada, H. Tanigawa, S. Kasai, N. Ohshima, S. Fukami, N. Ishiwata, Y. Nakatani, and T. Ono, "Control of Domain Wall Position by Electrical Current in Structured Co/Ni Wire with Perpendicular Magnetic Anisotropy," Appl. Phys. Express **1**, 101303 (2008).

33. A. Thiaville, S. Rohart, É. Jué, V. Cros, and A. Fert, "Dynamics of Dzyaloshinskii domain walls in ultrathin magnetic films," Europhys. Lett. **100**, 57002 (2012).

34. J. Cho, N.-H. Kim, S. Lee, J.-S. Kim, R. Lavrijsen, A. Solignac, Y. Yin, D.-S. Han, N. J. J. van Hoof, H. J. M. Swagten, B. Koopmans, and C.-Y. You, "Thickness dependence of the interfacial Dzyaloshinskii–Moriya interaction in inversion symmetry broken systems," Nat. Commun. **6**, 7635 (2015).

35. A. J. Huber, F. Keilmann, J. Wittborn, J. Aizpurua, and R. Hillenbrand, "Terahertz Near-Field Nanoscopy of Nanodevices," Nano Lett. **8**, 3766 (2008).

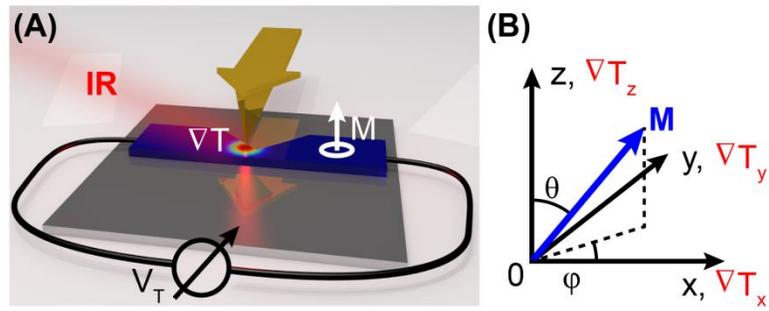

**Figure 1 (A)** Schematic of the measurement setup, showing AFM tip and 1 μm wide magnetic bar containing a central 500 nm triangularly shaped notch. The AFM tip is illuminated by IR radiation generating a sub-diffraction confined near-field at the tip apex. The power loss of the radiation on the sample surface leads to a nanoscale heat source, as indicated by the red spot. The thermal gradient $\nabla T$ generates an electromotive force, $V_T$, measured as a function of tip position. Near the boundaries of the nanostructure contributions to $V_T$ are expected due the magnetization, $M$, of the sample perpendicular to the plane via the anomalous Nernst effect (ANE). **(B)** Definition of the coordinate system.

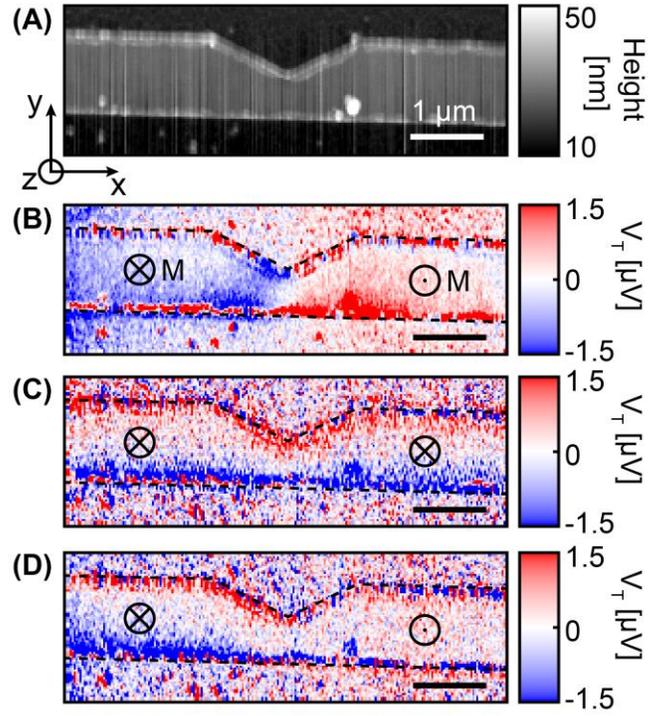

**Figure 2 (A)** Topographical image of the microbar. **(B)** Thermal EMF, $V_T$, as measured for oppositely magnetized domains with the domain wall located at the center of the constriction. The $V_T$ – map contains both, Seebeck and ANE contributions. **(C)** Difference map of $V_T(\theta = \pi) - V_T(\theta = 0)$ for single domain (homogeneously magnetized) bar. Since the Seebeck effect does not dependent on magnetization it is compensated, leading to a pure ANE contribution. **(D)** Difference map of $V_T(\theta = 0, \pi) - V_T(\theta = \pi)$ for the microbar containing a domain wall subtracted by the single domain case leaving the pure ANE contribution of the compartment to the left of the notch.

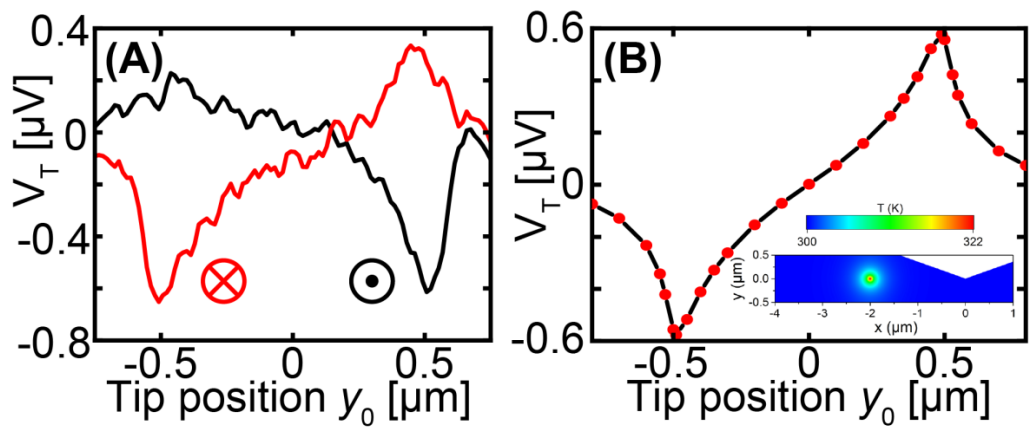

**Figure 3** (**A**) Thermal EMF, $V_T(y_0)$, averaged (12 lines with 15 nm separation) and smoothed (Savitzky-Golay) traces far from the notch with opposite magnetizations (indicated by the color). (**B**) Computed numerical line scan across the bar. The inset depicts the thermal gradient generated by the nanoscale heat source.